\documentclass[journal,10pt,onecolumn,draftclsnofoot,]{IEEEtran}
\ifCLASSINFOpdf
   \usepackage[pdftex]{graphicx}
   \graphicspath{{../pdf/}{../jpeg/}}
   \DeclareGraphicsExtensions{.pdf,.jpeg,.png}
\else
   \usepackage[dvips]{graphicx}
   \graphicspath{{../eps/}}
   \DeclareGraphicsExtensions{.eps}
\fi
\usepackage{epstopdf}
\usepackage{amsmath}
\usepackage{bm}
\usepackage{amsthm}
\UseRawInputEncoding
\usepackage{subfigure}
\usepackage{caption}
\usepackage{amssymb}
\usepackage{cite}
\usepackage{color}

\begin{document}
\captionsetup[figure]{font={small},labelformat={default},labelsep=period,name={Fig.},singlelinecheck=off}

\title{Secure Outage Analysis of RIS-Assisted Communications with Discrete Phase Control}

\author{Wei Shi, Jindan Xu, 
       Wei~Xu,~\IEEEmembership{Senior Member,~IEEE},
       Marco~Di~Renzo,~\IEEEmembership{Fellow,~IEEE},
       and Chunming Zhao 

\thanks{Copyright (c) 2015 IEEE. Personal use of this material is permitted. However, permission to use this material for any other purposes must be obtained from the IEEE by sending a request to pubs-permissions@ieee.org.}

\thanks{Wei Shi, Wei Xu, Jindan Xu, and Chunming Zhao are with the National Mobile Communications Research Laboratory, Southeast University, Nanjing 210096, China (e-mail:\{wshi, wxu, jdxu, cmzhao\}@seu.edu.cn).}

\thanks{Marco Di Renzo is with Universit\'e Paris-Saclay, CNRS, CentraleSup\'elec, Laboratoire des Signaux et Syst\`emes, 3 Rue Joliot-Curie, 91192 Gif-sur-Yvette, France (e-mail: marco.di-renzo@universite-paris-saclay.fr).} 

}

\maketitle

\begin{abstract}
This correspondence investigates a reconfigurable intelligent surface (RIS)-assisted wireless communication system with security threats. The RIS is deployed to enhance the secrecy outage probability (SOP) of the data sent to a legitimate user. By deriving the distributions of the received signal-to-noise-ratios (SNRs) at the legitimate user and the eavesdropper, we formulate, in a closed-form expression, a tight bound for the SOP under the constraint of discrete phase control at the RIS. The SOP is characterized as a function of the number of antenna elements, $N$, and the number of discrete phase choices, $2^b$. It is revealed that the performance loss in terms of SOP due to the discrete phase control is ignorable for large $N$ when $b\!\geq\!3$. In addition, we explicitly quantify this SOP loss when binary phase shifts with $b\!=\!1$ is utilized. It is identified that increasing the RIS antenna elements by $1.6$ times can achieve the same SOP with binary phase shifts as that by the RIS with ideally continuous phase shifts. Numerical simulations are conducted to verify the accuracy of these theoretical observations.

\end{abstract}

\IEEEpeerreviewmaketitle
\section{Introduction}

Reconfigurable intelligent surface (RIS) is a metasurface that consists of a large number of passive reflecting elements with integrated low power electronics \cite{1}\cite{2}. A main feature of an RIS is that the amplitude and phase of each reflecting element can be independently controlled through software, thereby realizing passive beamforming (BF) for improving the signal quality at intended receivers. Due to these merits, RISs have been considered for various wireless applications, e.g., in millimeter-wave (mmWave) \cite{3} and Terahertz (THz) \cite{4} communications, to enhance spectral and energy efficiencies~\cite{5}.

In recent years, physical layer security (PLS) has gained considerable interest for securing wireless communications. As a complement to conventional cryptographic methods, PLS ensures secure communications by exploiting the dynamics of propagation channels. Since RISs have the ability of adjusting the propagation channels, their deployment empowers the design of PLS with an additional dimension by exploiting passive BF. In order to maximize the theoretical secrecy, literature \cite{6,7,8} investigated joint optimization of the active and passive BFs at the transmitter and RIS, respectively. In \cite{9}, the average secrecy rate (SR) was characterized for an RIS-assisted two-way communication system through a lower bound. More recently in \cite{10}, the SR was further analyzed for a system where the RIS reflection is utilized as a multiplicative randomness against a wiretapper.

Besides the analysis of SR, secrecy outage probability (SOP) is another relevant performance measure to quantify the performance of PLS, especially for systems undergoing slow-varying channels. The SOP is defined as the probability that the instantaneous secrecy capacity falls below a target SR. The SOPs of RIS-aided communication systems have been investigated in \cite{11,12,13}. Both analytical and asymptotic analyses have been provided to reveal the impacts of key system parameters on SOP. In particular, the work in \cite{12} considered the SOP of an RIS-aided unmanned aerial vehicle (UAV) relay system. In \cite{13}, the authors analyzed the SOP as well as the probability of non-zero secrecy capacity of an RIS-aided device-to-device (D2D) communication system. However, most studies on SOP analysis considered RIS with continuous phase shifts, which leads to unaffordable high complexity in practice. Even though discrete phase shifts have been considered for RIS reflection optimizations, e.g., in \cite{14}\cite{15}, few studies have been conducted on theoretical performance analysis with discrete RIS phase shifts especially in terms of SOP. Quantitative insight on RIS design has only been discovered for some non-security scenarios \cite{16}. This is because discrete phase shifts make the performance expressions of the cascaded RIS channels much less tractable. In particular for secure communications, it can be even challenging to directly derive the corresponding cascaded channel distributions of both the legitimate user and eavesdropper.

In this work, we investigate the SOP of an RIS-assisted secure communication system and quantitatively characterize the impacts of discrete phase shifts in closed-form expressions. Concretely, we first derive the exact distributions of the received signal-to-noise-ratios (SNRs) at the legitimate user and the eavesdropper. Then, we present a closed-form expression for a tight upper bound of the SOP. Based on the obtained expressions, the asymptotic scaling law of SOP is characterized in high-SNR regimes. In particular, the SOP decreases with the slope of ${\rm e}^{-0.8N}$ for large $N$ and $b\!\geq\!3$, where $N$ is the number of RIS elements and $b$ is the number of quantization bits. Compared with the ideally continuous phase control, we further obtain that the performance loss in terms of SOP caused by using binary phase shifts, i.e., $b\!=\!1$, can be compensated by deploying a larger RIS with size $1.6N$.

The remainder of this paper is organized as follows. The system model is introduced in Section \uppercase\expandafter{\romannumeral2}. In Section \uppercase\expandafter{\romannumeral3}, we derive the distribution of the received SNRs at the legitimate user and at the eavesdropper. In Section \uppercase\expandafter{\romannumeral4}, we provide a closed-form expression for an upper bound of the SOP, and we study it in notable asymptotic regimes. Simulation results and conclusions are given in Section \uppercase\expandafter{\romannumeral5} and \uppercase\expandafter{\romannumeral6}, respectively.

\section{System Model}
\begin{figure}[!t]
    \setlength{\abovecaptionskip}{0pt}
    \setlength{\belowcaptionskip}{0pt}
    \centering
    \includegraphics[width=3.0in,height=4cm]{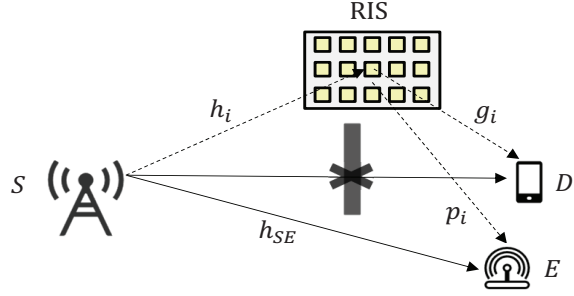}
    \caption{The system model of an RIS-assisted secure communication.}
    \label{fig1} \end{figure}

We consider an RIS-assisted secure communication system consisting of a source ($S$), an RIS with $N$ reflecting elements, a legitimate user ($D$), and an eavesdropper ($E$), as illustrated in Fig. \ref{fig1}. The direct link between $S$ and $D$ is assumed to be blocked by obstacles, such as buildings, which is likely to occur at high frequency bands. In this scenario, the data transmission between $S$ and $D$ is ensured by the RIS. The eavesdropper is at a location where it can overhear the information from both $S$ and the RIS.\footnote{The direct link between $S$ and $E$ exists when the eavesdropper is not blocked by obstacles \cite{11}\cite{13}.} Nodes $S$, $D$, and $E$ are equipped with a single antenna for transmission and reception and all links experience Rayleigh fading.\footnote{The RIS-related links can be modeled as Rayleigh fading, also as \cite{11}\cite{13}, when the RIS is not optimally deployed to ensure strong LoS links.} The channel coefficients of the $S$-RIS, RIS-$D$, RIS-$E$, and $S$-$E$ links are respectively denoted by $h_i$, $g_i$, $p_i$, and $h_{S\!E} \sim \mathcal{CN}(0,1)$, where ${\cal {CN}}$ is the complex Gaussian distribution. By applying channel estimation methods in \cite{3,4} and the references therein, we assume that the channel coefficients of $h_i$ and $g_i$ are perfectly known to $S$. However, the channel gains of $p_i$ and $h_{S\!E}$ are not available to $S$, as the eavesdropper is usually a passive device that does not emit signals.

By assuming quasi-static flat fading channels, the signal received at $D$ is expressed as
\begin{equation}
r_D=\sqrt P\left[{\eta(d_{S\!R}d_{R\!D})}^{-\upsilon/2}\sum_{i=1}^{N}{h_ig_i{\rm e}^{j\phi_i}}\right]x+n_D,
\label{equal1}
\end{equation}
where $P$ denotes the transmit power at $S$, $x$ is the transmit signal, $\eta\in(0,1]$ is the RIS amplitude reflection coefficient with $\eta=1$ corresponding to lossless reflection, $\left\{\phi_i\right\}_{i=1}^N$ represents the phase shift of the $i$th reflecting element of the RIS, and $n_D \sim \mathcal{CN}(0,N_0)$ is the additive white Gaussian noise (AWGN) with zero mean and variance $N_0$. Without loss of generality, the signal power is normalized, i.e., $\mathbb{E}[\left|x\right|^2]=1$ where $\mathbb{E}\left[\cdot\right]$ denotes the expectation of a random variable (RV). In addition, $d_{S\!R}$ and $d_{R\!D}$ are the distances of the $S$-RIS and RIS-$D$ links, respectively, and $\upsilon$ is the path loss exponent.

From (\ref{equal1}), the received SNR at $D$ is calculated as
\begin{equation}
\gamma_D=\frac{{\eta}^2P\left|\sum_{i=1}^{N}{h_ig_i{\rm e}^{j\phi_i}}\right|^2}{N_0d_{S\!R}^\upsilon d_{R\!D}^\upsilon}.
\label{equal2}
\end{equation}
In the case of ideally continuous phase shifts, $\gamma_D$ is maximized by setting the phases of the RIS elements equal to $\phi_i\!=\!\phi_i^{\rm opt}\!\triangleq\!-\angle\left({h_i}{g_i}\right)$, where $\angle$ returns the phase of the complex number. This optimized phase compensates the phase shift introduced by fading channels. Due to hardware limitations, however, the phase shifts $\left\{\phi_i\right\}_{i=1}^N$ of the RIS elements are usually limited to a finite number of controllable discrete values. In particular, the set of discrete phase shifts is denoted by $\mathcal{A}\triangleq\left\{0, \frac{2\pi}{2^b}, ..., \frac{(2^b-1)2\pi}{2^b}\right\}$, where $b$ is the number of quantization bits. In this case, usually, the phase shift of the $i$th RIS element, $\phi_i^{\rm sub}\in\mathcal{A}$, is chosen as  
\begin{equation}
\phi_i^{\rm sub}\triangleq\arg\mathop{\min}\limits_{{\phi\in\mathcal{A}}}\left\{\left|\phi_i^{\rm opt}-\phi\right|\right\}.
\label{equal3}
\end{equation}
Then, the received SNR in the presence of the discrete phase shifts is rewritten as
\begin{align}
\gamma_D\!=\!{\bar{\gamma}_{S\!R\!D}}\left|\sum_{i=1}^{N}{h_ig_i{\rm e}^{j\phi_i^{\rm sub}}}\right|^2,
\label{equal4}
\end{align}
where ${\bar{\gamma}}_{S\!R\!D}\triangleq\frac{{\eta}^2P}{N_0d_{S\!R}^\upsilon d_{R\!D}^\upsilon}$ denotes the average SNR.

The eavesdropper receives signals from the direct link from $S$ and the reflected link from the RIS. Then, the received signal at $E$ is written as
\begin{equation}
r_E\!=\!\sqrt P\!\left[\eta{(d_{S\!R}d_{R\!E})}^{-\upsilon/2}\sum_{i=1}^{N}{h_ip_i{\rm e}^{j\phi_i^{\rm sub}}}\!+\!d_{S\!E}^{-\upsilon/2}h_{S\!E}\right]\!x+n_E,
\label{equal5}
\end{equation}
where $d_{R\!E}$ and $d_{S\!E}$ denote the distances of the RIS-$E$ and $S$-$E$ links, respectively, and $n_E \sim \mathcal{CN}(0, N_0)$ is the AWGN at $E$. Then, the received SNR at $E$ is
\begin{align}
\gamma_E\!=\!\left|\sqrt{{\bar{\gamma}}_{S\!R\!E}}\sum_{i=1}^{N}{h_ip_i{\rm e}^{j\phi_i^{\rm sub}}}+\sqrt{{\bar{\gamma}}_{S\!E}}h_{S\!E}\right|^2,
\label{equal6}
\end{align}
where ${\bar{\gamma}}_{S\!R\!E}\triangleq\frac{{\eta}^2P}{N_0d_{S\!R}^\upsilon d_{R\!E}^\upsilon}$ and ${\bar{\gamma}}_{S\!E}\triangleq\frac{{\eta}^2P}{N_0d_{S\!E}^\upsilon}$ represent the average SNRs of the $S$-RIS-$E$ and $S$-$E$ links, respectively.

\section{Distributions of the Received SNRs}
In order to analyze the SOP of the system, we need to first characterize the distributions of $\gamma_D$ and $\gamma_E$.
\subsection{Distribution of $\gamma_D$} 
Let us denote the quantization error of phase shifts by $\Theta_i\!\triangleq\!\phi_i^{\rm sub}\!-\!\phi_i^{\rm opt}$, which is uniformly distributed \cite{14}\cite{15}\cite{17}, i.e., $\Theta_i\!\sim\!\mathcal {U}\left(-2^{-b}\pi,2^{-b}\pi\right)$. Then, $\gamma_D$ in (\ref{equal4}) is rewritten as 
\begin{align}
\gamma_D&\!\overset{({\rm a})}{=}\!{\bar{\gamma}_{S\!R\!D}}\left|\sum_{i=1}^{N}{h_ig_i{\rm e}^{j(\Theta_i+\phi_i^{\rm opt})}}\right|^2\!\overset{({\rm b})}{=}\!{\bar{\gamma}_{S\!R\!D}}\left|\sum_{i=1}^{N}{\left|h_i\right|\left|g_i\right|{\rm e}^{j\Theta_i}}\right|^2\nonumber\\
&\!\overset{({\rm c})}{=}\!\bar{\gamma}_{S\!R\!D}\left(X^2\!+\!Y^2\right)\!\overset{({\rm d})}{=}\!\gamma_{D_1}\!+\!\gamma_{D_2},
\label{equal7}
\end{align}
where $({\rm a})$ follows by the identity $\phi_i^{\rm sub}\!=\!\Theta_i+\phi_i^{\rm opt}$, $({\rm b})$ is obtained by using $\phi_i^{\rm opt}\!=\!-\angle\left({h_i}{g_i}\right)$, $({\rm c})$ follows by the definitions $X\!\triangleq\!\sum_{i=1}^{N}{\left|h_i\right|\left|g_i\right|\cos{\Theta_i}}$ and $Y\!\triangleq\!\sum_{i=1}^{N}{\left|h_i\right|\left|g_i\right|\sin{\Theta_i}}$, and $({\rm d})$ is obtained by defining $\gamma_{D_1}\!\triangleq\!{\bar{\gamma}}_{S\!R\!D}X^2$ and $\gamma_{D_2}\!\triangleq\!{\bar{\gamma}}_{S\!R\!D}Y^2$. 
Before deriving the distribution of $\gamma_D$, we introduce the following lemma.

\emph{Lemma~1:} If $N$ is large, $\gamma_{D_1}$ and $\gamma_{D_2}$ are statistically independent. Also, the cumulative distribution function (CDF) of $\gamma_{D_1}$ and the probability density function (PDF) of $\gamma_{D_2}$ are, respectively,
\begin{equation}
F_{\gamma_{D_1}}\left(x\right)=1-\frac{1}{2}{\rm erfc}\left(\frac{\alpha\!+\!\sqrt x}{\beta}\right)\!-\!\frac{1}{2}{\rm erfc}\left(\frac{\sqrt x\!-\!\alpha}{\beta}\right),
\label{equal8}
\end{equation}
\begin{equation}
f_{\gamma_{D_2}}\left(y\right)=\frac{\lambda^{\mu}y^{\mu-1}}{\Gamma\left(\mu\right)}{\rm e}^{-\lambda y},
\label{equal9}
\end{equation}
where $\alpha\!=\!m_1\sqrt{{\bar{\gamma}}_{S\!R\!D}}$, $\beta\!=\!\sqrt2\sigma_1\sqrt{{\bar{\gamma}}_{S\!R\!D}}$, $m_1\!=\!\frac{N\pi}{4}{\rm sinc}\left(2^{-b}\right)$, $\sigma_1^2\!=\!\frac{N}{2}[1+{\rm sinc}(2^{1-b})]-\frac{N\pi^2}{16}{\rm sinc}^2\left(2^{-b}\right)$, $\lambda\!=\!\frac{1}{2\sigma_2^2{\bar{\gamma}}_{S\!R\!D}}$, $\mu\!=\!\frac{1}{2}$, $\sigma_2^2\!=\!\frac{N}{2}[1-{\rm sinc}(2^{1-b})]$, where ${\rm sinc}(x)\!\triangleq\!\frac{\sin{\pi x}}{\pi x}$ and $\Gamma\left(\cdot\right)$ is the Gamma function [18, Eq.~(8.310)].

\emph{Proof:} The proof of the independence of $\gamma_{D_1}$ and $\gamma_{D_2}$ for large values of $N$ is provided in Appendix~A. Specifically, by applying the central limit theorem (CLT) \cite{19}, $X$ and $Y$ converge in distribution to Gaussian RVs for large $N$. Since $\left|h_i\right|$ and $\left|g_i\right|$ are independently distributed Rayleigh RVs with mean $\sqrt\pi/2$ and variance $(4-\pi)/4$, we obtain $\mathbb{E}\left[X\right]=m_1$, ${\rm Var}[X]=\sigma_1^2$, $\mathbb{E}\left[Y\right]=0$, and ${\rm Var}[Y]=\sigma_2^2$. It follows 
\begin{equation}
X\xrightarrow{\rm{d}}\mathcal{N}(m_1, \sigma_1^2),~Y\xrightarrow{\rm{d}}\mathcal{N}(0, \sigma_2^2),
\label{equal10}
\end{equation}
where $\xrightarrow{\rm{d}}$ denotes the convergence in distribution by virtue of the CLT. $\gamma_{D_1}$ is a non-central $\chi^2$ RV and $\gamma_{D_2}$ is a central $\chi^2$ RV with one degree of freedom, where $\chi^2$ denotes the Chi-square distribution. Then, by using [20, Eq.~(2.3-35)] and [21, Eq.~(27)], $F_{\gamma_{D_1}}\left(\cdot\right)$ is derived. The PDF $f_{\gamma_{D_2}}\left(\cdot\right)$ is obtained from [20, Eq.~(2.3-28)]. The proof completes.$\hfill\blacksquare$

By applying \emph{Lemma~1} and (\ref{equal7}), the CDF of $\gamma_D$ equals
\begin{align}
F_{\gamma_D}\left(z\right)=&\iint_{D}{f_{\gamma_{D_1},\gamma_{D_2}}\left(x,y\right){\rm{d}}x{\rm{d}}y}\nonumber\\
\overset{({\rm d})}{=}&\int_{0}^{z}{f_{\gamma_{D_2}}\left(y\right)F_{\gamma_{D_1}}\left(z-y\right){\rm{d}}y}\nonumber\\
\overset{({\rm e})}{=}&\int_{0}^{z}{\frac{\lambda^{\mu}y^{\mu-1}}{\Gamma\left(\mu\right)}e^{-\lambda y}}\times\left[1\!-\!\frac{1}{2}{\rm erfc}\left(\frac{\sqrt{z\!-\!y}\!+\!\alpha}{\beta}\right) \right. \nonumber\\
&\left.-\frac{1}{2}{\rm erfc}\left(\frac{\sqrt{z\!-\!y}\!-\!\alpha}{\beta}\right)\right]{\rm{d}}y,
\label{equal11}
\end{align}
where $D\triangleq\left\{\left(x,y\right):~x+y\leq z,~x>0,~y>0\right\}$, $({\rm d})$ utilizes the independence of $\gamma_{D_1}$ and $\gamma_{D_2}$ in \emph{Lemma~1}, and $({\rm e})$ is obtained by using (\ref{equal8}) and (\ref{equal9}).

\subsection{Distribution of $\gamma_E$} 
Let us first consider the distribution of $Z\!\triangleq\!\sqrt{{\bar{\gamma}}_{S\!R\!E}}\sum_{i=1}^{N}{h_ip_i{\rm e}^{j\phi_i^{\rm sub}}}\!+\!\sqrt{{\bar{\gamma}}_{S\!E}}h_{S\!E}$. Then, $\gamma_E$ in (\ref{equal6}) can be calculated according to the relationship of $\gamma_E=\left|Z\right|^2$.
The distribution of $Z$ is provided in the following lemma.

\emph{Lemma~2:} For large $N$, $Z\xrightarrow{\rm{d}}\mathcal{CN}(0, N{\bar{\gamma}}_{S\!R\!E}+{\bar{\gamma}}_{S\!E})$, and the real and imaginary parts of $Z$ are independent RVs with equal variance.

\emph{Proof:} See Appendix B.$\hfill\blacksquare$

As disclosed in \emph{Lemma~2}, $\gamma_E$ has an exponential distribution with mean $N{\bar{\gamma}}_{S\!R\!E}+{\bar{\gamma}}_{S\!E}$. Therefore, the PDF of $\gamma_E$ is
\begin{equation}
f_{\gamma_E}\left(x\right)=\epsilon {\rm e}^{-\epsilon x},~x\geq0
\label{equal12}
\end{equation}
where $\epsilon={1}/{(N{\bar{\gamma}}_{S\!R\!E}+{\bar{\gamma}}_{S\!E})}$.

\section{Theoretical Analysis of the SOP}
\subsection{SOP Analysis}
The SOP is an essential performance metric to quantify the performance of PLS, which is defined as the probability that the instantaneous secrecy capacity falls below a target positive SR $C_{\rm th}$. From \cite{11,12,13}\cite{22}, the SOP is calculated as
\begin{align}
SOP&={\rm Pr}\left(\ln{\left(1+\gamma_D\right)}-\ln{\left(1+\gamma_E\right)}<C_{\rm th}\right)\nonumber\\
&=\int_{0}^{\infty}{F_{\gamma_D}\left(\left(1+x\right)\varphi-1\right)}f_{\gamma_E}\left(x\right){\rm{d}}x,
\label{equal13}
\end{align}
where $\varphi\triangleq {\rm e}^{C_{\rm th}}$. It is still difficult to compute (\ref{equal13}) because the CDF of $\gamma_D$ in (\ref{equal11}) involves an intractable integral. Thus, instead of seeking for a closed-form expression for the SOP, we derive an upper bound as follows
\begin{equation}
SOP < \int_{0}^{\infty}{F_{\gamma_{D_1}}\left(\left(1+x\right)\varphi-1\right)}f_{\gamma_E}\left(x\right){\rm{d}}x = \overline {SOP}.
\label{equal14}
\end{equation}

\emph{Remark~1:} Note that $\overline {SOP}$ in (\ref{equal14}) is tight when $N$ is large, because $\gamma_{D_1}\!\!\gg\!\!\gamma_{D_2}$ holds with high probability, whose proof is provided in Appendix~C.

\emph{Lemma~3:} The upper bound in (\ref{equal14}) can be expressed as
\begin{align}
\overline {SOP}=1-\frac{1}{2}\left(I_1+I_2\right),
\label{equal15}
\end{align}
where $I_1$ and $I_2$ are respectively defined as follows
\begin{align}
I_1 =~&{\rm erfc}\left(\frac{\sqrt{\varphi\!-\!1}\!+\!\alpha}{\beta}\right)\!-\!\frac{2\sqrt{A}}{\beta}{\rm e}^{AB^2\!+\!\frac{\left(\varphi\!-\!1\right)\epsilon}{\varphi}\!-\!\frac{\alpha^2}{\beta^2}}\nonumber\\
&\times\left[1\!-\!{\rm erf}\left(B\sqrt A\!+\!\frac{\sqrt{\varphi\!-\!1}}{2\sqrt A}\right)\right],
\label{equal16}
\end{align}
\begin{align}
I_2 =~&{\rm erfc}\left(\frac{\sqrt{\varphi\!-\!1}\!-\!\alpha}{\beta}\right)\!-\!\frac{2\sqrt A}{\beta}{\rm e}^{AB^2\!+\!\frac{\left(\varphi\!-\!1\right)\epsilon}{\varphi}\!-\!\frac{\alpha^2}{\beta^2}}\nonumber\\
&\times\left[1\!-\!{\rm erf}\left(-B\sqrt A\!+\!\frac{\sqrt{\varphi\!-\!1}}{2\sqrt A}\right)\right],
\label{equal17}
\end{align}
where $A=\frac{\beta^2\varphi}{4(\beta^2\epsilon+\varphi)}$ and $B=\frac{2\alpha}{\beta^2}$.

\emph{Proof:} See Appendix D.$\hfill\blacksquare$

\subsection{Asymptotic SOP Analysis}
We consider application scenarios characterized by a low transmission rate but high security requirements, such as for the Internet of Things \cite{23}. The target SR $C_{\rm th}$ can be so small that $\varphi\rightarrow 1$. In this case, we obtain
\begin{align}
&\overline {SOP}\mathop\to\limits^{\left(\rm f\right)}\sqrt{\frac{1}{2\sigma_1^2{\bar{\gamma}}_{S\!R\!D}\epsilon\!+\!1}}{\rm e}^{-\frac{m_1^2{\bar{\gamma}}_{S\!R\!D}\epsilon}{2\sigma_1^2{{\bar{\gamma}}_{S\!R\!D}}\epsilon+1}}\mathop\to\limits^{\left(\rm g\right)} \sqrt{\frac{1}{2\sigma_1^2{\bar{\gamma}}_{S\!R\!D}\epsilon}}{\rm e}^{-\frac{m_1^2}{2\sigma_1^2}}\nonumber\\
=&\sqrt{\frac{1}{k\!\left[1\!+\!{\rm sinc}\left(2x\right)\!-\!\frac{\pi^2}{8}{\rm sinc}^2\left(x\right)\!\right]}} {\rm e}^{-\frac{1}{\frac{16\left[1+{\rm sinc}\left(2x\right)\right]}{\pi^2{\rm sinc}^2\left(x\right)}-2}N},
\label{equal18}
\end{align}
where (f) is obtained from (\ref{equal15}) by setting $\varphi\rightarrow1$, and (g) holds true in the high-SNR regime when  ${\bar{\gamma}}_{S\!R\!D}\gg\{{\bar{\gamma}}_{S\!R\!E}, {\bar{\gamma}}_{S\!E}\}$ and by using the following inequalities
\begin{align}
2\sigma_1^2{\bar{\gamma}}_{S\!R\!D}\epsilon=&\frac{1\!+\!{\rm sinc}\left(2x\right)\!-\!\frac{\pi^2}{8}{\rm sinc}^2\left(x\right)}{{\bar{\gamma}}_{S\!R\!E}\!+\!{\bar{\gamma}}_{S\!E}/N}{\bar{\gamma}}_{S\!R\!D}\nonumber\\
\geq&\frac{{\bar{\gamma}}_{S\!R\!D}}{2\left({\bar{\gamma}}_{S\!R\!E}\!+\!{\bar{\gamma}}_{S\!E}\right)}\gg1,
\label{equal19}
\end{align}
where $x\triangleq2^{-b}$ and the inequality in (\ref{equal19}) follows by $1+{\rm sinc}\left(2x\right)\!-\!\frac{\pi^2}{8}{\rm sinc}^2\left(x\right)\geq\frac{1}{2}$ for $x\in\left(0, \frac{1}{2}\right]$. The last equality in (\ref{equal18}) is obtained by defining $k\!\triangleq\!{{\bar{\gamma}}_{S\!R\!D}}/\left({{\bar{\gamma}}_{S\!R\!E}}+\frac{1}{N}{\bar{\gamma}}_{S\!E}\right)$. 

By direct inspection of (\ref{equal18}), we evince that $\overline {SOP}$ decreases if ${\bar{\gamma}}_{S\!R\!D}$ increases, which means that enhancing the average SNR at the legitimate user always improves the secrecy performance even for a limited number of discrete phase shift status. In particular, we have $\overline {SOP}\rightarrow0$ when ${\bar{\gamma}}_{S\!R\!D}\rightarrow\infty$ while keeping $N$, $b$, ${\bar{\gamma}}_{S\!R\!E}$, and ${\bar{\gamma}}_{S\!E}$ fixed.

Besides, inspired by \cite{24}, we have the following remark to show the impact of the RIS location on SOP.

\emph{Remark~2:} The SOP improves when $\{d_{S\!R}, d_{R\!D}\}$ decreases and $d_{R\!E}$ increases. For large $N$, it is further found that the asymptotic SOP hardly changes with a moderate variation of $d_{S\!R}$. This behavior is also explained from the fact that reducing the distance from the source to RIS increases the received SNRs for both the legitimate user and the eavesdropper. Therefore, in the case that the location of the eavesdropper is unknown, we should prioritize deploying the RIS closer to the legitimate user than to the source.

\emph{Proof:} From (\ref{equal18}), we see that the location of the RIS is only reflected in the parameter $k=\frac{d_{R\!D}^{-\upsilon}}{d_{R\!E}^{-\upsilon}+\frac{1}{N}\left(\frac{d_{S\!E}}{d_{S\!R}}\right)^{-\upsilon}}$, and the asymptotic SOP in (\ref{equal18}) decreases as $k$ grows. $\hfill\blacksquare$

Moreover, the high-SNR expression in (\ref{equal18}) is also useful for understanding the asymptotic secrecy performance given the number of control bits of discrete phase shifts. Some relevant case studies are reported as follows.

\textbf{Case~1:} Under the assumption of continuous-valued phase shifts, i.e., $b=+\infty$, the SOP in (\ref{equal18}) reduces to
\begin{equation}
{\overline{SOP}}\bigg|_{b=+\infty}\rightarrow\sqrt{\frac{8}{k\left(16-{\pi^2}\right)}}{\rm e}^{-\frac{\pi^2}{32-2\pi^2}N}.
\label{equal20}
\end{equation}

\textbf{Case~2:} Under the assumption of $1$-bit binary phase shifts, i.e., $b=1$, the SOP in (\ref{equal18}) reduces to
\begin{equation}
{\overline{SOP}}\bigg|_{b=1}\rightarrow\sqrt{\frac{2}{k}}{\rm e}^{-\frac{N}{2}}.
\label{equal21}
\end{equation}

\textbf{Case~3:} Under the assumption of $b\!\geq\!3$, we prove in Appendix~E that the quantization noise, which is due to the use of discrete phase shifts, is one order-of-magnitude smaller than the SOP of the continuous phase shifts disclosed in Case~1.

\emph{Remark~3:} According to Case~1 and Case~2, the SOP tends to ${\overline{SOP}}|_{b=+\infty}\!\rightarrow\! \sqrt{{8{{\bar{\gamma}}_{S\!R\!E}}}/\left[{\left(16-{\pi^2}\right){{\bar{\gamma}}_{S\!R\!D}}}\right]}{\rm e}^{-0.8N}$ and ${\overline{SOP}}|_{b=1}\!\rightarrow\! \sqrt{{2{{\bar{\gamma}}_{S\!R\!E}}}/{{{\bar{\gamma}}_{S\!R\!D}}}}{\rm e}^{-0.5N}$ for sufficiently large $N$. We see that the SOP loss due to the $1$-bit quantization, compared to the ideal continuous-valued phase shifts, can be asymptotically compensated by increasing the number of RIS elements by about $1.6$ times.

\emph{Proof:} Let $N_1$ and $N_2$ denote the numbers of RIS elements corresponding to $b=+\infty$ and $b=1$, respectively. By solving ${\overline{SOP}}|_{b=1}\le{\overline{SOP}}|_{b=+\infty}$, it follows that $N_2\geq \frac{\pi^2}{16-\pi^2}N_1-{\rm ln}\frac{4}{16-\pi^2}\approx 1.6 N_1$. This completes the proof. $\hfill\blacksquare$

\section{Numerical Results}
In this section, Monte-Carlo simulations are illustrated to validate our analysis. The tested parameters are set to ${\bar{\gamma}}_{S\!E}\!=\!-5~\rm{dB}$ and ${\bar{\gamma}}_{S\!R\!E}\!=\!0~\rm {dB}$. 
\begin{figure}[!t]
    \setlength{\abovecaptionskip}{0pt}
    \setlength{\belowcaptionskip}{0pt}
    \centering
    \includegraphics[width=90 mm,height=6.5cm]{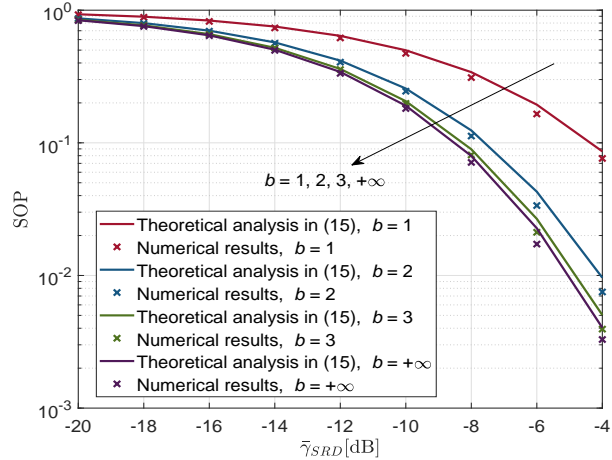}
    \caption{SOP versus ${\bar{\gamma}}_{S\!R\!D}$ for different values of $b$.}
    \label{fig2} \end{figure}
\begin{figure}[!t]
    \setlength{\abovecaptionskip}{0pt}
    \setlength{\belowcaptionskip}{0pt}
    \centering
    \includegraphics[width=90 mm,height=6.5cm]{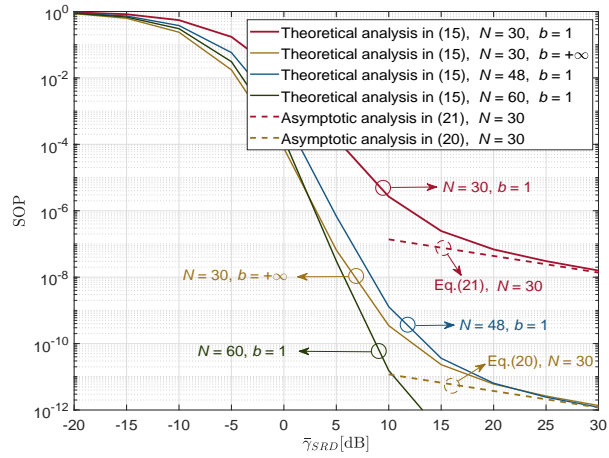}
    \caption{SOP versus ${\bar{\gamma}}_{S\!R\!D}$ for different values of $N$.}
    \label{fig3} \end{figure}
    
Fig. \ref{fig2} shows the impact of $b$ on the SOP when $N=30$ and $C_{\rm th}=0.05$. We observe that the analytical results in (\ref{equal15}) match well with the numerical curves. Monte-Carlo simulations are illustrated for values of the SOP no smaller than $10^{-3}$ due to the limited number of channel realizations simulated. As stated in Case~3, the gap between $b=3$ and $b=+\infty$ in Fig. \ref{fig2} does appear negligible for high SNRs.

In Fig. \ref{fig3}, we consider a larger $C_{\rm th}=0.2$ to verify the effectiveness of the asymptotic analysis. We plot the SOP for $N=30$ by setting $b=1$ and $b=+\infty$. As expected, the asymptotic expressions in (\ref{equal20}) and (\ref{equal21}) are quite tight in the high-SNR regime. Then, we plot the SOP for $b=1$ by setting $N=48$ (i.e., which is equal to $1.6 \times 30$) and $N=60$ ($2 \times 30$). We see that the setup $N=48$ with $b=1$ provides, in the high-SNR regime, the same SOP as the setup $N=30$ with $b=+\infty$, which validates the obtained guideline in \emph{Remark~3}.

\section{Conclusion}
This paper studied the SOP of an RIS-assisted communication system with discrete phase shifts. The main contribution is to unveil the achievable scaling law of SOP with respect to $N$ and $b$. Specifically, the increased number of RIS elements was quantified to compensate the performance loss caused by binary phase shifts.

\begin{appendices}
\section{Proof of The Independence of $\gamma_{D_1}$ and $\gamma_{D_2}$}
By taking into account that the distribution of $\Theta_i$ is symmetric around its mean value, which is equal to zero, we have $\mathbb{E}\left[XY\right]=0$ \cite{25}. Then, the covariance of $X$ and $Y$ is
\begin{equation}
{\rm Cov}\left[X,Y\right]=\mathbb{E}\left[XY\right]-\mathbb{E}\left[X\right]E\left[Y\right]=\mathbb{E}\left[XY\right]=0,
\label{equal22}
\end{equation}
which indicates that $X$ and $Y$ are uncorrelated RVs. 

For large values of $N$, $\sum_{i=1}^{N}{\left|h_i\right|\left|g_i\right|{\rm e}^{j\Theta_i}}$ is approximately a complex Gaussian RV by virtue of the CLT, and its real part $X$ and imaginary part $Y$ are jointly Gaussian RVs. Since two uncorrelated Gaussian RVs are independent as well, it follows that $\gamma_{D_1}$ and $\gamma_{D_2}$ are independent.

\section{Proof of Lemma 2}
First, we note that $h_i$ and $\phi_i^{\rm sub}$ are dependent RVs since $\phi_i^{\rm sub}$ and $\phi_i^{\rm opt}=-\angle\left({h_i}{g_i}\right)$ are correlated RVs. Since $\phi _i^{{\rm{sub}}}={\Theta _i}+\phi _i^{{\rm{opt}}}$, the RV $Z$ can be rewritten as 
\begin{align}
Z\!=\!\sqrt{{\bar{\gamma}}_{S\!R\!E}}\sum_{i=1}^{N}{\left|h_i\right|\left|p_i\right|{\rm e}^{j\left({\angle p_i}-{\angle g_i}+\Theta_i\right)}}\!+\!\sqrt{{\bar{\gamma}}_{S\!E}}h_{S\!E},
\label{equal23}
\end{align}
where ${\angle p_i}$ and ${\angle g_i}$ are uniformly distributed in $\left[0,2\pi\right)$. The RV $\Theta_{i}$ depends on the phase error at the RIS, and ${\angle p_i}$ and ${\angle g_i}$ depend on the positions of the eavesdropper and the legitimate user, respectively. Since the RV ${p_i} = \left| {{p_i}} \right|{{\rm e}^{j\angle {p_i}}}$ is independent of the three RVs $\left| {{h_i}} \right|$, $\angle {g_i}$, and ${\Theta _i}$ and has zero mean, we obtain $\mathbb{E}\left[\left|h_i\right|\left|p_i\right|{\rm e}^{j\left({\angle p_i}-{\angle g_i}+\Theta_i\right)}\right]\!=\!0$, and ${\rm Var}\left[\left|h_i\right|\left|p_i\right|{\rm e}^{j\left({\angle p_i}-{\angle g_i}+\Theta_i\right)}\right]\!=\!1$. For large values of $N$, $\sum_{i=1}^{N}{\left|h_i\right|\left|p_i\right|{\rm e}^{j\left({\angle p_i}-{\angle g_i}+\Theta_i\right)}}\xrightarrow{d}\mathcal{CN}(0, N)$ by virtue of the CLT, and then $Z\xrightarrow{d}\mathcal{CN}(0, N{\bar{\gamma}}_{S\!R\!E}+{\bar{\gamma}}_{S\!E})$.

Since $p_i$ is a circularly-symmetric Gaussian RV, we have ${\rm Pr}({\rm e}^{j\angle p_i})\!=\!{\rm Pr}({\rm e}^{j \angle p_i}{\rm e}^{j(-\angle g_i+\Theta_i)})$. Thus, $\left| {{p_i}} \right|{\rm e}^{j\left(\angle p_i-\angle g_i+\Theta_i\right)}$ is a zero-mean circularly-symmetric complex Gaussian RV as well. This implies that the real and imaginary parts of $Z$ are uncorrelated and hence independent since they are Gaussian distributed. Also, they have zero means and the same variance. 

\section{Proof of Remark 1}
Since the RV ${\gamma_{D_2}}/{\gamma_{D_1}}$ is nonnegative, we obtain
\begin{align}
{\rm Pr}\!\left(\frac{\gamma_{D_2}}{\gamma_{D_1}}\!<\!0.1\right)\mathop\geq\limits^{\left(\rm h\right)} 1-\frac{\mathbb{E}\left[\frac{\gamma_{D_2}}{\gamma_{D_1}}\right]}{0.1}\mathop=\limits^{\left(\rm i\right)}1-\frac{\mathbb{E}\left[{\gamma_{D_2}}\right]}{0.1\mathbb{E}\left[{\gamma_{D_1}}\right]},
\label{equal24}
\end{align}
\hspace*{-0.2cm} where (h) is obtained by applying the Markov inequality \cite{19}, and (i) comes from the fact that $\gamma_{D_1}$ and $\gamma_{D_2}$ are independent.

Furthermore, $\frac{\mathbb{E}\left[{\gamma_{D_2}}\right]}{\mathbb{E}\left[{\gamma_{D_1}}\right]}$ is calculated as
\begin{align}
\frac{\mathbb{E}\left[{\gamma_{D_2}}\right]}{\mathbb{E}\left[{\gamma_{D_1}}\right]}\!=\!\frac{\sigma_2^2}{m_1^2\!+\!\sigma_1^2}\!=\!\frac{8[1-{\rm sinc}(2x)]}{{(N\!-\!1)\pi^2}{\rm sinc}^2\left(x\right)\!+\!8[1\!+\!{\rm sinc}(2x)]},
\label{equal25}
\end{align}
\hspace*{-0.14cm}where $x=2^{-b}\in(0,\frac{1}{2}]$ and it is easy to check that $\frac{\mathbb{E}\left[{\gamma_{D_2}}\right]}{\mathbb{E}\left[{\gamma_{D_1}}\right]}$ is increasing as $x$ grows by calculating its first order derivative. It follows
\begin{align}
\frac{\mathbb{E}\left[{\gamma_{D_2}}\right]}{\mathbb{E}\left[{\gamma_{D_1}}\right]}\leq \left.\frac{\mathbb{E}\left[{\gamma_{D_2}}\right]}{\mathbb{E}\left[{\gamma_{D_1}}\right]}\right|_{x=\frac{1}{2}}=\frac{2}{N+1}.
\label{equal26}
\end{align}

Therefore, (\ref{equal24}) is further expressed as
\begin{align}
{\rm Pr}\!\left(\frac{\gamma_{D_2}}{\gamma_{D_1}}\!<\!0.1\right)\geq 1-\frac{\mathbb{E}\left[{\gamma_{D_2}}\right]}{0.1\mathbb{E}\left[{\gamma_{D_1}}\right]}=1-\frac{20}{N+1}.
\label{equal27}
\end{align}

When $N$ is large, we finally obtain ${\rm Pr}\!\left(\frac{\gamma_{D_2}}{\gamma_{D_1}}\!<\!0.1\right)\!\mathop\to\!1$, which means that $\gamma_{D_1}\!\gg\!\gamma_{D_2}$ holds with high probability.

\section{Proof of Lemma 3}
By inserting (\ref{equal8}) and (\ref{equal12}) in (\ref{equal14}), the upper bound of the SOP can be rewritten as
\begin{align}
\overline {SOP}=&\int_{0}^{\infty}\left[1-\frac{1}{2}{\rm erfc}\left(\frac{\alpha+\sqrt{\left(1+x\right)\varphi-1}}{\beta}\right)\right.\nonumber\\
&-\left.\frac{1}{2}{\rm erfc}\left(\frac{\sqrt{\left(1+x\right)\varphi-1}-\alpha}{\beta}\right)\right]\epsilon {\rm e}^{-\epsilon x}{\rm{d}}x\nonumber\\
=&1-\frac{1}{2}\left(\int_{0}^{\infty}{\rm erfc}\left(\frac{\alpha+\sqrt{\left(1+x\right)\varphi-1}}{\beta}\right)\epsilon {\rm e}^{-\epsilon x}{\rm{d}}x\right.\nonumber\\
&+\left.\int_{0}^{\infty}{\rm erfc}\left(\frac{\sqrt{\left(1+x\right)\varphi-1}-\alpha}{\beta}\right)\epsilon {\rm e}^{-\epsilon x}{\rm{d}}x\right)\nonumber\\
=&1-\frac{1}{2}\left(I_1+I_2\right).
\label{equal28}
\end{align}
The integral $I_1$ is calculated by using the integration by parts method. We have
\begin{align}
I_1~=~&\int_{0}^{\infty}{\rm erfc}\left(\frac{\sqrt{\left(1+x\right)\varphi-1}+\alpha}{\beta}\right)\epsilon {\rm e}^{-\epsilon x}{\rm{d}}x\nonumber\\
=~&\left.-{\rm erfc}\left(\frac{\sqrt{\left(1+x\right)\varphi-1}+\alpha}{\beta}\right){\rm e}^{-\epsilon x}\right|_{x=0}^\infty\nonumber\\
&+\int_{0}^{\infty}{\rm e}^{-\epsilon x}{\rm{d}}\left({\rm erfc}\left(\frac{\sqrt{\left(1+x\right)\varphi-1}+\alpha}{\beta}\right)\right)\nonumber\\
=~&{\rm erfc}\left(\frac{\sqrt{\varphi\!-\!1}\!+\!\alpha}{\beta}\right)\!-\!\frac{1}{\sqrt\pi\beta}J_1,
\label{equal29}
\end{align}
where $J_1$ is expressed as 
\begin{align}
J_1&=\int_{0}^{\infty}{\varphi {\rm e}^{-\epsilon x}\frac{{\rm e}^{-{\left(\sqrt{\left(1+x\right)\varphi-1}+\alpha\right)^2}/{\beta^2}}\ }{\sqrt{\left(1+x\right)\varphi-1}}}{\rm{d}}x\nonumber\\
&=2{\rm e}^{\frac{\left(\varphi-1\right)\epsilon}{\varphi}-\frac{\alpha^2}{\beta^2}}\int_{\sqrt{\varphi-1}}^{\infty}{{\rm e}^{-\left(\frac{1}{\beta^2}+\frac{\epsilon}{\varphi}\right)t^2-\frac{2\alpha}{\beta^2}t}}{\rm{d}}t\nonumber\\
&=2{\rm e}^{\frac{\left(\varphi-1\right)\epsilon}{\varphi}-\frac{\alpha^2}{\beta^2}}\sqrt{\pi A}{\rm e}^{AB^2}\left[1\!-\!{\rm erf}\left(B\sqrt A\!+\!\frac{\sqrt{\varphi-1}}{2\sqrt A}\right)\right],
\label{equal30}
\end{align}
$t=\sqrt{\left(1+x\right)\varphi-1}$, and the last equality is obtained by using [18, Eq.~(3.322)], where $A=\frac{\beta^2\varphi}{4(\beta^2\epsilon+\varphi)}$ and $B=\frac{2\alpha}{\beta^2}$. Further, by substituting (\ref{equal30}) into (\ref{equal29}), $I_1$ is obtained as shown in (\ref{equal16}). Analogously, $I_2$ is calculated by replacing $\alpha$ in (\ref{equal16}) with $-\alpha$, which yields the desired result in (\ref{equal17}).

\section{Proof of Case 3}
We apply the Taylor expansion to the SOP in (\ref{equal18}). More precisely, around $x=0$ and for small values of $x=2^{-b}<1$, we have
\begin{equation}
\sqrt{\frac{1}{k\!\left[1\!+\!{\rm sinc}\left(2x\right)\!-\!\frac{\pi^2}{8}{\rm sinc}^2\left(x\right)\!\right]}}={\rm c}_1+{\rm c}_2x^2+O\left(x^4\right),
\label{equal31}
\end{equation}
\begin{equation}
{\rm e}^{-\frac{1}{\frac{16\left[1+{\rm sinc}(2x)\right]}{\pi^2{\rm sinc}(x)}-2}N}={\rm e}^{-{\rm c}_3N}+O\left(x^4\right),
\label{equal32}
\end{equation}
where ${\rm c}_1=\sqrt{\frac{8}{k\left(16-\pi^2\right)}}$, ${\rm c}_2=\frac{\pi^2}{6}{\rm c}_1$, and ${\rm c}_3=\frac{\pi^2}{32-2\pi^2}$. 

By substituting (\ref{equal31}) and (\ref{equal32}) in (\ref{equal18}), the SOP is further rewritten as 
\begin{align}
\overline {SOP}&\rightarrow\left({\rm c}_1+{\rm c}_2x^2+O\left(x^4\right)\right)\left({\rm e}^{-{\rm c}_3N}+O\left(x^4\right)\right)\nonumber\\
&= {\rm c}_1{\rm e}^{-{\rm c}_3N}+{\rm c}_2{\rm e}^{-{\rm c}_3N}x^2+O\left(x^4\right).
\label{equal33}
\end{align}

Since the first term in (\ref{equal33}) is the SOP for $b\!=\!+\infty$ as given in Case~1, the performance loss due to finite values of $b$ is dominated by the second term in (\ref{equal33}), i.e., ${\rm c}_2{\rm e}^{-{\rm c}_3N}x^2$. If we restrict that the performance loss is one order of magnitude smaller than Case~1 for $b\!=\!+\infty$, it gives $\left({{\rm c}_2{\rm e}^{-{\rm c}_3N}x^2}\right)/\left({{\rm c}_1{\rm e}^{-{\rm c}_3N}}\right)<{1}/{10}$, which implies $x<\sqrt{3/\left(5{\pi}^2\right)}$ and equivalently $b\geq3$.

\end{appendices}

\end{document}